\documentclass[12pt]{iopart}
\usepackage{graphicx} 

\begin{document}
\title[A hybrid double-dot in silicon]{A hybrid double-dot in silicon}

\author{M. Fernando Gonzalez-Zalba\footnote{Contributed equally to this work}, Dominik Heiss\footnote{Contributed equally to this work} and Andrew J. Ferguson}
\address{ Microelectronics, Cavendish Laboratory, Cambridge University, Cambridge, UK }
\ead{ajf1006@cam.ac.uk}

\begin{abstract}
We report electrical measurements of a single arsenic dopant atom in the tunnel-barrier of a silicon SET. As well as performing electrical characterization of the individual dopant, we study series electrical transport through the dopant and SET. We measure the triple points of this hybrid double dot, using simulations to support our results, and show that we can tune the electrostatic coupling between the two sub-systems.
\end{abstract}

\maketitle

The study of single dopants in silicon is motivated by the prospect of quantum computation with long-lived electronic and nuclear spins \cite{Kane1998}. The observation of individual dopant states in nanoscale field effect transistors has been important progress towards this goal. Electrical transport spectroscopy has enabled positive identification of dopants \cite{Sellier2006, Ono2007, Tan2010} as well as investigations of their energy level structure in the presence of an interface \cite{Lansbergen2008, Pierre2010}. More recently, spin-read out of an dopant electronic state has been performed using a silicon single electron transistor (SET) to sense the occupancy of a nearby dopant  \cite{Morello2010}. The ability to measure the dopant spin state is important for future experiments that probe the electron and nuclear spin coherence of single dopants.

A parallel direction in silicon based quantum computation has been the development of few electron quantum dots, where the spin state of single confined electrons (or electron pairs) is of interest. This follows the progress made in the GaAs material system, but with the advantage of a reduced nuclear spin environment. Specifically, in double quantum dots the well-established mechanism of spin-blockade enables the singlet and triplet states to be distinguished \cite{Ono2002, Johnson2005}. In GaAs, this has enabled experiments on gate defined few electron quantum dots allowing the investigation of spin lifetime, spin coherence and exchange interaction between electrons in the two dots \cite{Hanson2007, Petta2005}. In silicon, spin blockade has been observed in double quantum dots \cite{Shaji2008, Liu2008, Borselli2011}, and spin measurements performed in single quantum dots \cite{Xiao2010, Hayes2009, Simmons2011}.

In this article, we report the electrical characterization of double dot formed from a single arsenic atom and a silicon SET. This approach combines the research on dopants and quantum dots and could provide a new way to read out the long-lived spin state of a dopant using spin blockade. In contrast to an earlier study \cite{Golovach2011}, our SET is gate defined allowing electrostatic control over both the dopant and SET, and consequent observation and analysis of the triple points.

The device fabrication starts by growth of a 10 nm sacrificial oxide on a high-resistivity ($>$7000 $\Omega$cm) (100) silicon wafer. Ohmic contacts are defined by optical lithography and ion implantation of phosphorus (15~keV, $10^{15}$~cm$^{-2}$) and dopants included by low-dose (15~keV, $10^{11}$~cm$^{-2}$) ion-implantation of arsenic.  The sacrificial silicon oxide is removed after the implant and a 10~nm SiO$_2$ gate oxide is regrown at 850$^\circ$C, which also anneals out the implantation damage. We perform a forming gas anneal at 450$^\circ$C for 30~min followed by a rapid thermal anneal for 15~s at 1050$^\circ$C to reduce the interface trap and fixed oxide charge density.  The As profile was calculated by an implantation Monte-Carlo simulator \cite{SRIM} and has a maximum at 10~nm from the interface and a density of 4$\times$10$^{16}$cm$^{-3}$. By comparison, the residual phosphorous doping is estimated to be smaller than 10$^{12}$~cm$^{-2}$. We note that a larger As density at the interface is expected due to segregation to the interface during the thermal processing \cite{Grove1964}.

\begin{figure}[H]
	\centering
		\includegraphics{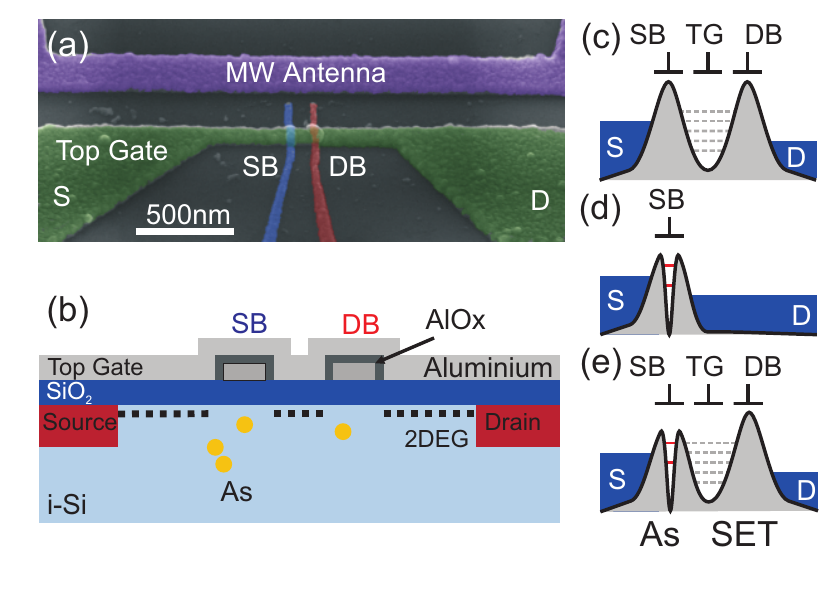}
	\caption{(a) Scanning electron microscopy image of an identical device. An on-chip coplanar stripline, for electron spin resonance, is included next to the sample but is not used in this experiment. (b) Schematic representation of the sample cross-section. (c-e) Schematic representation of the modes of device operation: (c) single electron transistor,(d) single dopant device, (e) dopant-SET hybrid device.}
	\label{Fig1}
\end{figure}

Subsequent to the silicon processing, surface gates are fabricated by electron beam lithography and thermal evaporation of aluminium. \Fref{Fig1} shows a scanning electron microscopy image of an identical device and its schematic cross-section. In a first step, two gates 40~nm wide and 100~nm apart are defined by evaporation of a 25~nm thick layer of aluminium.  They are used to form the tuneable source and drain tunnel barriers. After thermal oxidation at 150$^\circ$C for 5~min creating 5~nm aluminium oxide, a second electrically-independent 60~nm thick top-gate is deposited over the barriers. This top-gate defines the 120~nm wide channel of the SET and leads that overlap with the doped ohmic contacts. Prior to measurement the samples undergo a nitrogen ambient post-fabrication anneal at 350$^\circ$C for 15~min. The interface trap density, measured on simultaneously processed field effect transistors, by means of low frequency split C-V method \cite{Kommen1973}, is 1.8$\times$10$^{10}$cm$^{-2}$.

The device can be operated in one of the three different modes depicted in \fref{Fig1}: SET (c), single dopant (d), and hybrid dopant-SET device (e). For SET operation, the top-gate is set well above threshold and the tunnel barriers are biased to locally deplete the electron accumulation layer underneath. This forms an isolated island of electrons between the barriers leading to Coulomb oscillations in charge transport \cite{Angus2007}.  The second mode allows transport spectroscopy of individual As impurity atoms in the silicon substrate. The top-gate and drain tunnel barrier are biased well above threshold allowing the study of subthreshold phenomena underneath the source tunnel barrier. Finally, the third mode of operation permits the formation of a tuneable capacitively coupled dopant-SET hybrid. Here, the SET is defined while one of the barriers is tuned in resonance with a dopant transition.

Electrical transport measurements are performed at the base temperature of a dilution refrigerator (electron temperature of 200~mK) using radio-frequency reflectometry \cite{Schoelkopf1998}.  This technique probes the reflection coefficient of a resonant circuit that includes the device as a circuit element. As the impedance, in our case the differential conductance, of the device changes so does the reflection coefficient of the resonant circuit. This technique allows an increase of bandwidth over a standard dc or lock-in measurement. The sample was embedded in a rf-tank circuit formed by a surface mount 390~nH inductor and a parasitic capacitance (500~fF) to ground.  An rf-carrier signal is applied to the source of the device at the resonant frequency of 360~MHz and the cryo-amplified reflected signal is homodyne detected \cite{Angus2008}. A bias tee on the sample board permits the simultaneous measurement of the two terminal dc conductance.

\begin{figure}[H]
	\centering
		\includegraphics{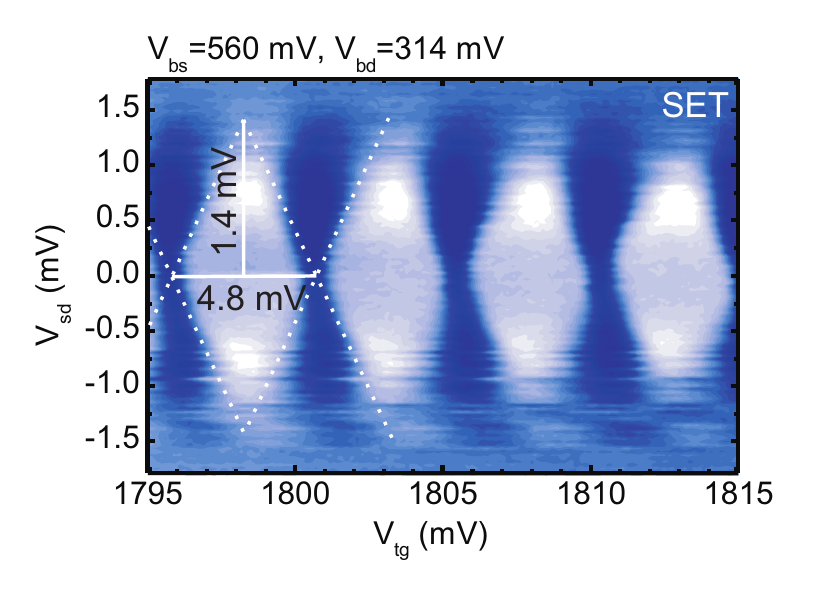}
	\caption{Measurement of the device in SET operation mode using rf-reflectometry. Coulomb diamonds with a gate period of 4.8~mV and a charging energy of 1.4~meV are observed.}
	\label{Fig2}
\end{figure}

To form the SET tunnel barriers, a bias of 560~mV and 314~mV is applied to the source and drain barriers respectively.  Periodic Coulomb diamonds are observed over a large range of top-gate bias (\fref{Fig2}), similar to earlier studies on undoped devices \cite{Angus2007}. From the diamonds we extract a voltage period of $\Delta$V$_{tg}$=4.8~mV and a charging energy of E$_c^{SET}$=1.4$\pm$0.1~meV, which corresponds to $\alpha_{SET}=C_{tg}/C_{SET}$=0.29, where $C_{tg}$ is the capacitive coupling of the topgate to the SET and $C_{SET}$ is the total SET capacitance. Due to our gate geometry, where the top-gate controls electron density in the island as well as the leads, we are not able to deplete the SET to the few electron limit. However, few electron quantum dots have been measured in a similar geometry but with separated gates controlling the leads and the island \cite{Lim2009b,Xiao2010b, Lim2011} .

\begin{figure}[H]
	\centering
		\includegraphics{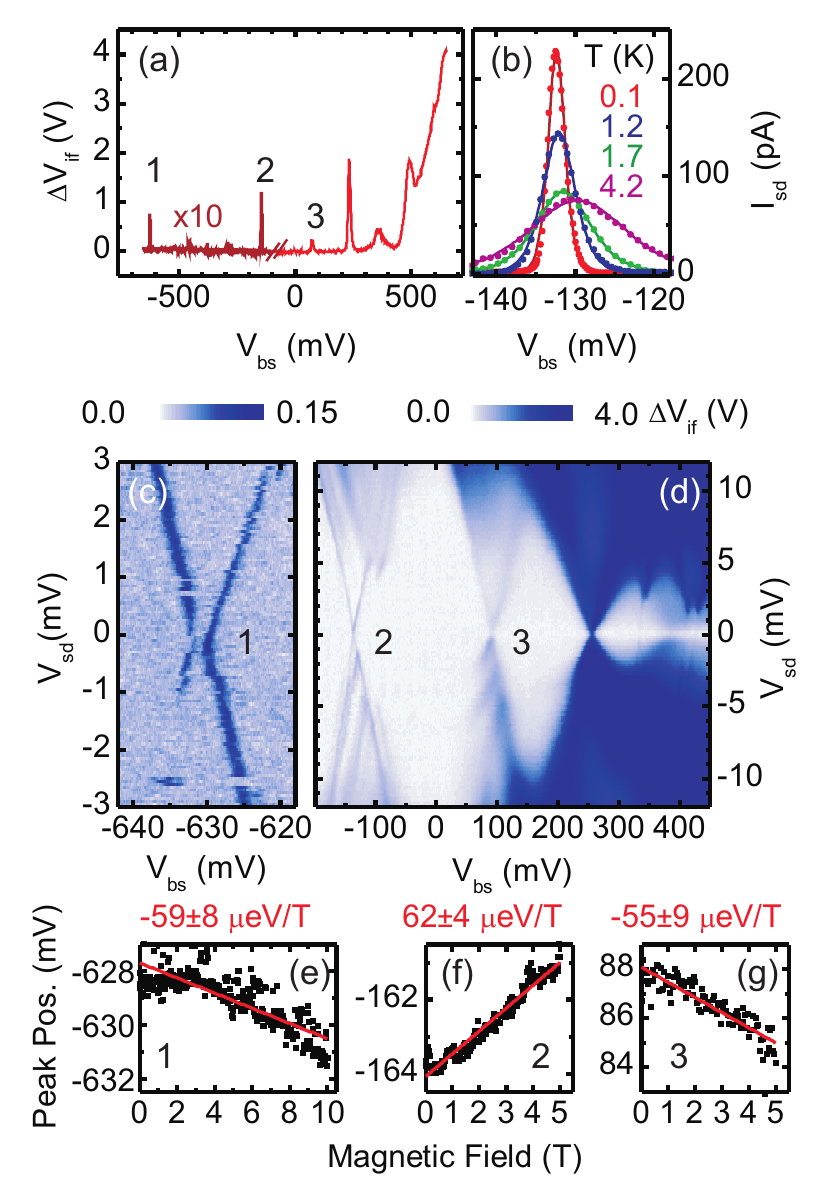}
	\caption{(a) Turn-off characteristics of the drain tunnelling barrier. Subthreshhold features labelled 1,2 and 3 are observed. (b) Tunnelling current of transition 2 as a function of temperature. In this measurement V$_{sd}$=-0.1~mV.  (c,d) Coulomb diamonds of the subthreshold features. (e-g) Magnetic field dependence of the peak position of transition 1,2 and 3, respectively.}
	\label{Fig3}
\end{figure}

We now describe electrical transport in the sub-threshold region beneath a single barrier, where we focus on the source barrier. The top-gate and drain barrier are set above threshold (V$_{tg}$=1.94~V), while the rf-response is measured as a function of V$_{bs}$ (\fref{Fig3}(a)) Below the conduction band edge (V$_{bs}$=430~mV), the data show electrical transport through states labeled 1, 2, 3 in \fref{Fig3}. These features appear at the same bias voltages in several cool down cycles. We identify these states as dopants in the barrier due to their charging energies, as extracted from (\fref{Fig3} (c,d)), being larger than 10~meV. In contrast, when we measure undoped samples, parasitic quantum dots are formed in the channel due to disorder at the interface, and these have a charging energy below 5~meV \cite{Podd2010}. To further investigate the nature of these states, we measured the line shape of the tunnel current as a function of temperature (symbols in \fref{Fig3}(b)). This is fitted to the expected behaviour for resonant tunnelling through a discrete state (lines in \fref{Fig3}(b)). In particular the maximum current increases with decreasing temperature, in contrast to Coulomb blockade through a multi-level system such as the SET. This is consistent with transport though a dopant with well separated energy levels ($\Delta E>k_B T$). Therefore, we attribute the transitions labelled 1, 2 and 3 to Arsenic dopants, which typically show charging energies on the order of 29-35~meV close to the Si/SiO$_2$ interface \cite{Sellier2006,Lansbergen2008}. 

We next examine the coupling, given by $\alpha=$C$_{bs}/$C$_\Sigma$, of the barrier gate to the different dopant transitions. This coupling is a direct translation of V$_{bs}$ to energy change on the dopant site. The obtained values were $\alpha_{1}$=0.21 , $\alpha_{2}$=0.10, $\alpha_{3}$=0.08. We expect the electrostatic characteristics of the device to change when changing the biasing conditions from V$_{bs}$=430~mV to -630~mV. The capacitive coupling of a dopant to the gate electrode ($C_{bs}$) and the contacts (C$_s$,C$_d$) is inherently changed due to screening effects and changes of the effective barrier width as a function of V$_{bs}$. As a result, it is not possible to identify which transitions stem from the same dopant site by comparing these values alone.

We perform magneto-spectroscopy on these states, applying an in-plane magnetic field perpendicular to the current direction (\fref{Fig3} (e-g)). The peak position is converted to a chemical potential shift using $\alpha$, and is in agreement with the expected Zeeman shift of 58$~\mu$eV/T for an electron with g-factor=2. Transitions 1 and 3 shift to lower energies corresponding to transport through the lower Zeeman sub-level. This behaviour is expected for tunnelling through an ionized donor (D$^+$-D$^0$ transition) \cite{Sellier2006,Lansbergen2008}. In contrast, transition 2 shifts to higher energies, suggesting the lower Zeeman level is already occupied and tunnelling takes place through an already occupied (neutral) donor (D$^0$-D$^-$ transition). Accordingly, we identify the three sub-threshold peaks as follows: 1 corresponds to D$^+$-D$^0$ transition of an As donor; 2 to the corresponding D$^0$-D$^-$ transition; and 3 is identified as the D$^+$-D$^0$ transition of an additional As donor. For typical donor charging energies of 29-35~meV we expect the corresponding D$^0$-D$^-$ transition to be around V$_{bs}$=400~mV, where several conductance peaks are observed that cannot be clearly identified.

Following this, the charging energy for transition 1 and 2 is determined as $E_c$=73~meV. Here, we use a average value of $\alpha_{1,2}$=0.155 and bias voltage difference $\Delta$V$_{bs}$=470~mV. Such an increased charging energy has not been observed so far, but may be a consequence our relatively large dopant density, the presence of another ionized As atom leading to an increased ionization energy \cite{Koiller2006, Rahman2010}. Another possibility is that, an orbital Stark shift \cite{Rahman2009} increases the observed charging energy. The electric field at the dopant site varies for the two transitions, due to the difference in biasing conditions. Additionally, a weaker screening effect induced by the metallic dominated interface is expected for the charged impurity atom in comparison to the neutral dopant \cite{Rahman2011}.  

Investigation of the drain barrier revealed two additional resonant tunnelling features consistent with transport through a neutral donor. As in the case of transition 3 the corresponding D$^0$-D$^-$ transition could not be clearly identified. Consequently, these states were not investigated further.

We now turn our attention to the hybrid dopant-SET system, investigating the sequential transport through a single dopant and the SET. A positive voltage of V$_{tg}$=2~V is applied to the top-gate while the source barrier gate voltage is set to V$_{bd}$=375~mV. The drain barrier is tuned such that transport is governed by tunnelling through transition 2, since in this configuration the strongest dopant-SET is observed.

\begin{figure}[H]
	\centering
		\includegraphics{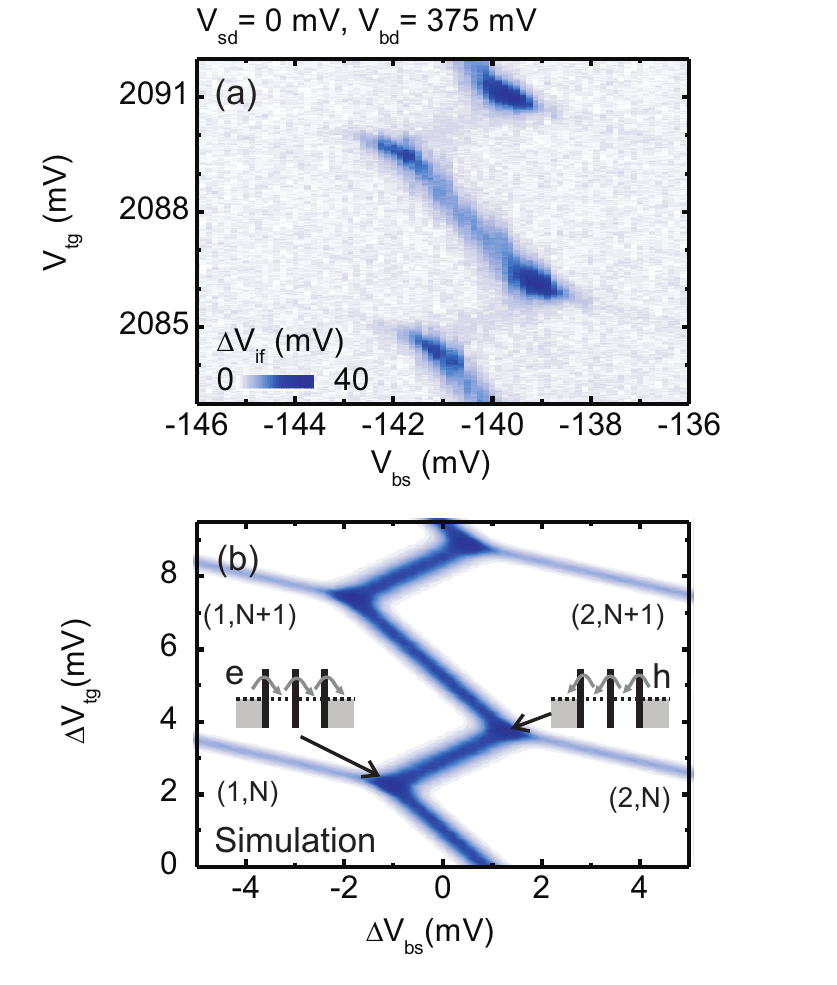}
	\caption{(a) Rf-reflectometry measurement as a function of top-gate bias V$_{tg}$ and source barrier bias V$_{bs}$. Four triple points can be observed within this range. (b) Simulation of the measurement presented in (a). The charge state is labelled in the form (n, m), where n denotes the number of electrons on the dopant site, while m stands for the electron number on the SET. The inset illustrates the difference between electron and hole transport process.}
	\label{Fig4}
\end{figure}

Again using rf-reflectometry, we measured the hybrid system formed by transition 2 and the SET, as a function of top-gate voltage V$_{tg}$ and barrier V$_{bs}$ (\fref{Fig4}(a)). As in a double quantum dot, electrical conduction occurs at points with three-fold degeneracy of the charge state, known as triple points \cite{Wiel2002}. Transport through the system can be described by a capacitance model, and we simulated the differential conductance to compare to the rf-reflectometry measurement (\fref{Fig4}(b)). The software SIMON \cite{SIMON} was used to calculate tunnelling probabilities by Monte-Carlo simulation. The SET was described as a metallic island with a constant density of states. The dopant was simulated as a discrete energy level system represented by delta functions. An excitation voltage of 125~$\mu$V, corresponding to -97~dBm of rf-power at the tank circuit, was used to obtain the differential conductance.

For increasing top-gate voltage we observe the charge transitions of the SET with a gate period of $\Delta$V$_{tg}$=4.8~mV in agreement with the characterization of the SET alone (\fref{Fig4}(b)). The dopant transition has been identified as D$^0$-D$^-$, therefore, the electron number increases from 1 to 2 for increasing barrier voltage V$_{bs}$. Weak lines of increased conductance are observable at the charge transitions due to elastic cotunnelling processes \cite{DeFranceschi2001, Liu2005}.

\begin{figure}[H]
	\centering
		\includegraphics{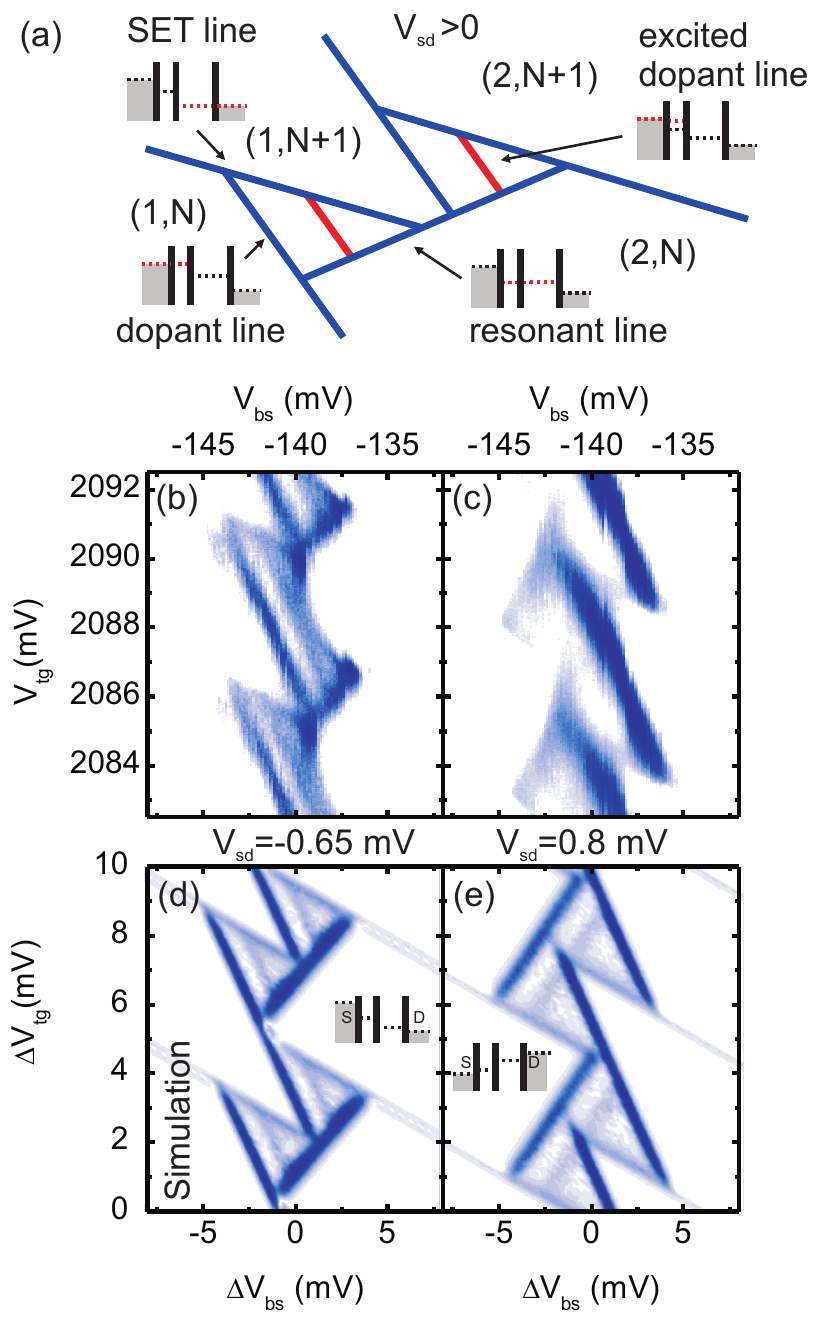}
	\caption{(a) Schematic representation of the bias triangles observed for V$_{sd}>0$. The resonant conditions defining the borders of the triangles are indicated. (b,c) Rf-reflectometry measurement of the hybrid system for V$_{sd}$=-0.65 and 0.8~mV, respectively. (d,e) Simulation of the measurement presented in (b,c). The SET is described by the capacitive coupling to drain C$_d$=74~aF, the top-gate C$_{tg}$=34~aF and a total capacitance C$_{SET}$=116~aF. For the dopant we find the source (C$_s$=0.7~aF), barrier (C$_{bs}$=0.34) and total (C$_{As}$=2.1~aF) capacitance.}
	\label{Fig5}
\end{figure}

In the finite bias regime $|$V$_{sd}|>$0 the conductance regions change from triple-points to triangles (\fref{Fig5}(a)). We measure (\fref{Fig5}(b,c)) and simulate (\fref{Fig5}(d,e)) these bias-triangles for both bias polarities. The dimensions of the bias triangles are related to the bias voltage through the corresponding values of $\alpha$, $\alpha_{bs}=e|$V$_{sd}|/\delta$V$_{bs}$, $\alpha_{tg}=e|$V$_{sd}|/\delta$V$_{tg}$. From the measurement presented in \fref{Fig5}(b,c), we can extract $\alpha_{bs}$=0.16 and $\alpha_{tg}$=0.28 for the dopant site and the SET, respectively. Despite the change in biasing, this is in reasonable agreement with the values obtained from isolated measurement of the dopant ($\alpha_{2}$=0.10) and the SET ($\alpha_{SET}$=0.29) stability diagram.

An additional feature parallel to the dopant line is visible within the bias triangle (\fref{Fig5}(b)), at an energy of 0.27$\pm$ 0.03~meV from the ground state. Such lines of increased differential conductance within the bias triangle can arise from modulation of density of states in the leads \cite{Fuechsle2010}. Alternatively, such features can be observed when the energy level of an excited dopant state enters the bias window. We cannot distinguish the origin of this line, but note that in the 2e configuration of a dopant a valley-spin excited state around 1~meV has been observed \cite{Lansbergen2011}. We do not detect the additional line for V$_{sd}$=0.8~meV, which is a consequence of the asymmetry in the hybrid system.

\begin{figure}[H]
	\centering
		\includegraphics{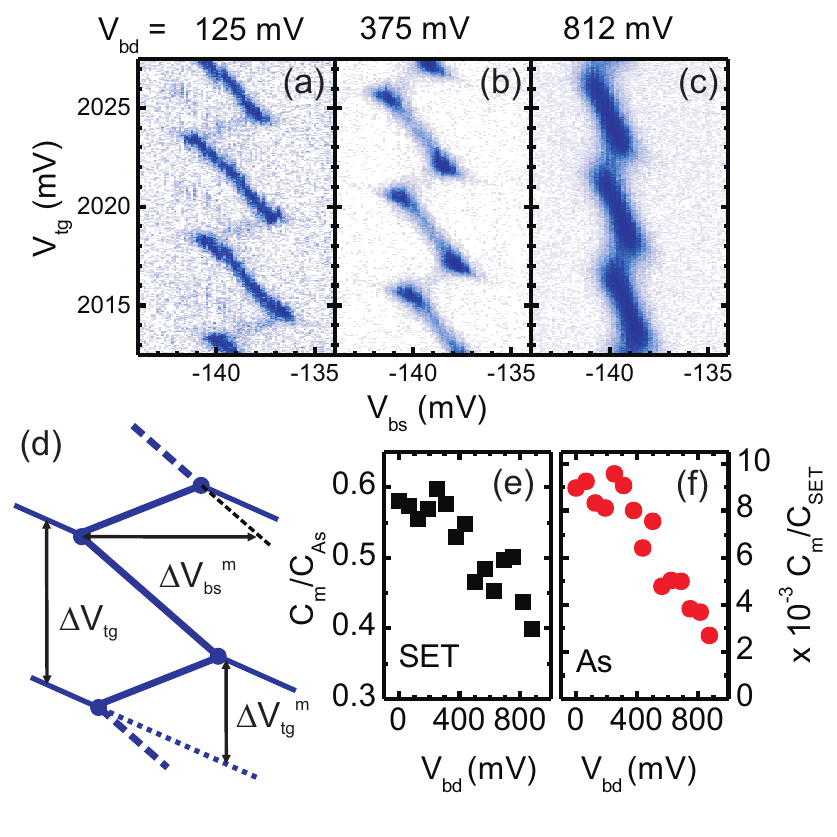}
	\caption{(a-c) The triple points of the hybrid system for drain barrier voltages V$_{bd}$=125, 375, 812~mV, respectively. (d) Schematic representation of the triple point measurement indicating the voltage separation $\Delta$V$_{bs}^{m}$, $\Delta$V$_{tg}$ and $\Delta$V$_{tg}^{m}$ . (e,f) Electrostatic coupling $C_m/C_{As}$ and $C_m/C_{SET}$ as a function of drain barrier bias V$_{bd}$.   }
	\label{Fig6}
\end{figure}

We will now discuss the coupling of the dopant and the SET. A measure for the electrostatic coupling of the SET to the dopant is the ratio $C_m/C_{SET}=\Delta$V$_{bs}^{m}/\Delta$V$_{bs}$ \cite{Yamahata2008}, where $C_m$ is the mutual capacitance between SET and dopant. The induced voltage change of the dopant line at the triple points $\Delta$V$_{bs}^{m}$ can be extracted from the measurement, as indicated in \fref{Fig6}(d) and $\Delta$V$_{bs}$=470~mV is the separation of the D$^+$-D$^0$ and the D$^0$-D$^-$ transition of the dopant. The data in \fref{Fig4}(a) shows $\Delta$V$_{bs}^{m}$=2.3~mV, resulting in $C_m/C_{SET}$=0.008. In analogy the electrostatic effect of the dopant on the SET is $C_m/C_{As}=\Delta$V$_{tg}^{m}/\Delta$V$_{tg}=0.53$. Using the previously determined total capacitance of the dopant C$_{As}$=2.1~aF and the SET C$_{SET}$=116~aF a mutual capacity of $C_m$=1.0$\pm$0.2~aF is estimated. Under these conditions the maximum source-drain current measured at the triple point for V$_{sd}$=-0.65~mV is 130~pA, which corresponds to a tunnelling time of 1.2~ns between dopant and SET.

Furthermore, the dopant-SET coupling can be tuned by changing the bias on the drain tunnel barrier $V_{bd}$ (\fref{Fig6}(a-c)). In \fref{Fig6}(e,f) the electrostatic coupling $C_m/C_{SET}$ and $C_m/C_{As}$ are plotted as a function of $V_{bd}$. As the voltage on the drain tunnel barrier is increased, the separation of the triple points decreases, which corresponds to a reduced electrostatic coupling between the dopant and the SET. This change arises partly from a reduced capacitive coupling $C_{m}$ as the SET island extends further away from the dopant site, partly from an increased $C_{SET}$ as the SET is coupled more strongly to the drain lead.     

To conclude, we have studied a hybrid double-dot formed by coupling an dopant and an SET in series. In transport spectroscopy, we observe triple points and bias triangles characteristic of a double quantum dot. The analogy with the double quantum dot could be taken further by reducing the SET to the few electron limit, which will allow spin-blockade to be observed. This will provide a new tool for the investigation of single electron spins on dopants in silicon.

\ack
We are grateful for support of EPSRC grant EP/H016872/1 and the Hitachi Cambridge Laboratory. AJF was supported by the Hitachi Research Fellowship. We acknowledge discussions with David Williams. MFGZ was supported by Obra Social La Caixa fellowship. 

\section*{References}
\bibliographystyle{unsrt}
\bibliography{Hybrid}

\end{document}